\documentclass[twocolumn]{aastex7}

\usepackage{amsmath}
\usepackage{multirow}
\usepackage{graphicx}
\usepackage{xcolor}
\usepackage{hyperref}

\newif\ifrevisionblue
\revisionbluetrue
\DeclareRobustCommand{\rev}[1]{{\ifrevisionblue\color{black}\fi #1}}
\newenvironment{revblock}{\begingroup\ifrevisionblue\color{black}\fi}{\endgroup}

\renewcommand{\DH}{D_{\rm H}}
\newcommand{\DM}{D_{\rm M}}
\newcommand{\DV}{D_{\rm V}}
\newcommand{\rd}{r_{\rm d}}
\newcommand{\OmL}{{\Omega}_{\rm M}^{\Lambda}}
\newcommand{\Om}{{\Omega}_{\rm M}}

\begin{document}

\title{A New Consistency Test for the $\Lambda$CDM Model using Radial and Transverse BAO Measurements}

\author{Xing-Han A. Zhao}
\affiliation{The International Division of the Second High School Attached to Beijing Normal University, Beijing 100192, P.R.China}
\email{xinghan.adam.zhao@gmail.com}

\author{Zheng Cai}
\affiliation{Department of Astronomy, Tsinghua University, Beijing 100084, P.R.China}
\email[show]{\url{zcai@tsinghua.edu.cn}}
\correspondingauthor{Zheng Cai}

\begin{abstract}
We present a calibration-free consistency test of spatially flat $\Lambda$CDM based on baryon acoustic oscillation (BAO)
distance measurements. The method forms ratios of BAO distances including the Hubble distance, the comoving angular
diameter distance, and the volume-averaged distance, so that the sound horizon scale cancels, and then
maps each observed ratio to an effective flat-$\Lambda$CDM matter density parameter, ${\Omega}_{\rm M}^{\Lambda}$,  defined as the value of
$\Omega_{\rm M}$ that reproduces the measured ratio within $\Lambda$CDM. Flat $\Lambda$CDM predicts that ${\Omega}_{\rm M}^{\Lambda}$ should be
independent of redshift and of the particular ratio used. For ratios involving the integrated distances, we
associate them with well-defined effective line-of-sight redshift intervals based on the
integral mean value theorem. We apply the test to BAO measurements from the Dark Energy Spectroscopic Instrument (DESI)
Data Release~1 and Data Release~2, propagating the full published BAO covariance matrices
into all derived ratios and ${\Omega}_{\rm M}^{\Lambda}$ constraints. \rev{We find no statistically significant deviation from flat $\Lambda$CDM in either DESI DR1 or DESI DR2.}
Within current uncertainties, the inferred ${\Omega}_{\rm M}^{\Lambda}$ values are broadly
consistent with a redshift-independent constant, providing an internal consistency check of flat $\Lambda$CDM that can be
strengthened straightforwardly as BAO measurements improve.
\end{abstract}

\keywords{cosmological parameters --- large-scale structure of Universe --- baryon acoustic oscillations}

\section{Introduction}
\label{sec:intro}

The discovery that the cosmic expansion is accelerating \citep{Riess1998,Perlmutter1999}
has made the physical origin of dark energy one of the central problems in modern cosmology.
In the standard picture, cosmic acceleration is attributed to a cosmological constant (vacuum energy)
\citep{Weinberg1989,Carroll2001} or to a dynamical component \citep{Ratra1988,Caldwell1998}.
At the same time, the spatially-flat $\Lambda$CDM model has achieved remarkable empirical success in describing
a broad set of observations, including the temperature anisotropies of the cosmic microwave background (CMB),
from the first detection by COBE to subsequent measurements of the acoustic peak structure
\citep{Smoot1992,Bennett1996,deBernardis2000,Halverson2002},
as well as the observed clustering and growth of large-scale structure \citep{Peebles1980}.
It therefore remains essential to develop robust, low-assumption consistency tests that can expose subtle departures
from flat $\Lambda$CDM or reveal residual observational systematics.

Baryon acoustic oscillations (BAO) provide one of the most robust late-time geometric probes.
The BAO feature was first robustly detected in galaxy redshift surveys in 2005 using the Sloan Digital Sky Survey
and the Two-degree Field Galaxy Redshift Survey (2dFGRS) \citep{Eisenstein2005,Cole2005},
and it has since become a standard ruler exploited by many surveys \citep{Alam2021}.
Modern spectroscopic surveys now deliver high-precision BAO constraints over a broad redshift range,
including recent measurements from the Dark Energy Spectroscopic Instrument (DESI)
\citep{DESI:2024uvr,DESI:2025zgx}.
A practical complication is that BAO are primarily measured as \emph{relative} distances in units of the sound horizon
at the drag epoch, and different BAO statistics constrain different combinations of line-of-sight and transverse distances.
This can make it non-trivial to apply popular $H(z)$-based null diagnostics to the full set of BAO observables
in a unified way.

\rev{This observational structure is precisely why BAO are well suited to a ratio-based null test. Anisotropic BAO measurements provide separate constraints on the radial and transverse distances, while isotropic BAO measurements constrain the angle-averaged distance; all are reported in units of the same ruler, $\rd$. Taking ratios therefore removes the absolute standard-ruler calibration and focuses the test on the shape of the distance--redshift relation. The approach also allows radial, transverse, and angle-averaged BAO information from the same survey to be compared within one framework, using the published covariance matrices rather than introducing an external $H_0$ or sound-horizon prior.}

A widely-used diagnostic for testing flat $\Lambda$CDM is the $Om$ statistic \citep{Sahni2008,Zunckel2008,Sahni2014,Shafieloo2012},
which is constructed to be constant in redshift for a flat $\Lambda$CDM cosmology and to vary when the expansion history deviates
from this model.
In practice, however, transverse BAO constraints involve integrated distance information, so straightforward applications of $Om$
naturally favor local expansion-rate measurements or require additional steps and assumptions to incorporate transverse BAO consistently.

In this work we devise a simple, ratio-based BAO consistency test that can simultaneously exploit radial and transverse BAO information
while remaining insensitive to the absolute BAO scale.
The basic idea is to form appropriately chosen ratios of BAO distance measurements so that the sound horizon (and overall calibration)
cancels, and then ask whether these ratios are mutually consistent with a single flat $\Lambda$CDM matter density parameter.
Each observed ratio can be mapped to an effective $\Lambda$CDM matter density, $\OmL$, and the redshift-independence of $\OmL$
across different ratios and redshift pairs provides a direct null test of flat $\Lambda$CDM.
To incorporate transverse and isotropic BAO information in a transparent manner, we use a redshift-matching strategy based on the
integral mean value theorem \citep{MVT} to relate integrated BAO distances to an effective line-of-sight distance over a well-defined
redshift interval.

This paper is organized as follows.
In Section~\ref{sec:method} we describe the construction of the ratio diagnostics, the mapping to $\OmL$, the redshift-matching procedure,
and the data sets used.
Section~\ref{sec:results} presents the results for DESI DR1 and DR2 BAO measurements.
We summarize and discuss implications in Section~\ref{sec:conclusion}, and provide additional details of the redshift-matching procedure
in an appendix.

\section{Methodology and Data}
\label{sec:method}
\label{sec:methodology}

\subsection{From the $Om$ diagnostic to BAO ratio tests}

The $Om$ diagnostic \citep{Sahni2008,Zunckel2008} is a widely-used null test of flat $\Lambda$CDM based on the expansion rate,
\begin{equation}
    Om(z) \equiv \frac{h^2(z)-1}{(1+z)^3-1}, \qquad h(z)\equiv H(z)/H_0,
\end{equation}
which is constant and equal to the present-day matter density parameter $\Om$ in a spatially flat $\Lambda$CDM universe.
To reduce sensitivity to the absolute calibration of $H_0$, a two-point version was proposed \citep{Sahni2014},
\begin{equation}\label{eq:2ptOm}
    Om(z_i,z_j)=\frac{h^2(z_i)-h^2(z_j)}{(1+z_i)^3-(1+z_j)^3}.
\end{equation}
While these diagnostics are naturally suited to local measurements of $H(z)$, modern BAO analyses provide a broader set of
observables, including transverse (integrated) distances. We therefore construct ratio-based diagnostics from BAO distance
measurements that (i) cancel the sound horizon scale $\rd$ and (ii) can be applied uniformly to radial and transverse BAO.

We adopt the standard BAO distance definitions:
the Hubble distance $\DH(z)\equiv c/H(z)$,
the comoving angular diameter distance
\begin{equation}\label{eq:DM}
    \DM(z)=\int_0^z \DH(z')\,{\rm d}z',
\end{equation}
and the volume-averaged BAO distance
\begin{equation}\label{eq:DV}
    \DV(z)=\left[z\,\DM^2(z)\,\DH(z)\right]^{1/3}.
\end{equation}
BAO constraints are commonly reported as $\DH/\rd$, $\DM/\rd$, or $\DV/\rd$, so appropriately chosen ratios of BAO distances
are independent of the absolute BAO scale $\rd$ (and, for our purposes, also independent of the overall normalization set by $H_0$).

\rev{The practical advantage for BAO data is threefold. First, the ratio construction uses the native BAO observables, so no reconstruction of an absolute $H(z)$ scale is required. Second, the same formalism can be applied to purely radial, mixed radial--transverse, and mixed radial--isotropic measurements. Third, because the inputs are the reported DESI distance vectors, the full published covariance matrix can be propagated directly into the derived ratios. In this sense the test is an internal BAO consistency check: it asks whether the relative distances measured by BAO can all be described by a single flat-$\Lambda$CDM matter density.}

\subsection{Ratio diagnostics and the effective $\Lambda$CDM matter density $\OmL$}

Our null test is based on mapping each observed distance ratio to an \emph{effective} matter density parameter in flat $\Lambda$CDM,
denoted by $\OmL$, defined as the value of $\Omega_{\rm M}$ that reproduces the observed ratio within a flat $\Lambda$CDM prediction.
If the Universe is described by flat $\Lambda$CDM, all ratios (and all redshift pairs) should yield a consistent, redshift-independent
$\OmL=\Om$. Any statistically significant redshift dependence of $\OmL$ indicates an inconsistency with flat $\Lambda$CDM (which could arise
from dynamical dark energy, spatial curvature, modified gravity, or residual observational systematics).

\rev{Here and below, $z_i$ and $z_j$ denote the DESI effective redshifts of the BAO redshift bins being compared. The indices $i$ and $j$ are therefore bin labels, not powers; when an ordered pair is used we take $z_i<z_j$ unless the ratio definition explicitly states otherwise.}

\subsubsection{Radial--radial test ($R_{\rm HH}$)}
We first construct a purely radial ratio using $\DH$ at two redshifts:
\begin{equation}\label{eq:RHH}
   \frac{\DH(z_i)}{\DH(z_j)}\equiv R_{\rm HH}(z_i,z_j)
   \xrightarrow{\Lambda}
   R_{\rm HH}^{\Lambda}(z_i,z_j)\equiv \frac{\DH^{\Lambda}(z_i)}{\DH^{\Lambda}(z_j)}.
\end{equation}
In a spatially flat $\Lambda$CDM model parameterized by $\OmL$, the Hubble distance is
\begin{equation}\label{eq:DHLCDM}
    \DH^{\Lambda}(z)=\frac{c}{H_0\,\sqrt{\OmL(1+z)^3+(1-\OmL)}}.
\end{equation}
Equating $R_{\rm HH}(z_i,z_j)=R_{\rm HH}^{\Lambda}(z_i,z_j)$ yields an analytic expression for $\OmL$ in terms of the observed ratio:
\begin{equation}\label{eq:OmL}
      \OmL = \rev{\frac{1-R_{\rm HH}^2(z_i,z_j)}{R_{\rm HH}^2(z_i,z_j)\,f(z_i)-f(z_j)}},
\end{equation}
where $f(z)\equiv(1+z)^3-1$. Note that $c$ and $H_0$ cancel out in the ratio and therefore do not affect the inferred $\OmL$.
\rev{With this convention, a noiseless flat-$\Lambda$CDM input returns the same value of $\OmL$ for every radial pair. A departure from flat $\Lambda$CDM would appear as a pair-dependent value of $\OmL$, or equivalently as a non-zero residual $R_{\rm HH}-R_{\rm HH}^{\Lambda}$ for any single reference value of $\Omega_{\rm M}$.}

\subsubsection{Radial--transverse test ($R_{\rm HM}$; diagonal)}
\label{sec:RHM}

We adopt the ``diagonal'' radial--transverse ratio,
\begin{equation}\label{eq:RHM}
   R_{\rm HM}(z)\equiv \frac{\DH(z)}{\DM(z)}
   \xrightarrow{\Lambda}
   R_{\rm HM}^{\Lambda}(z;\OmL)\equiv \frac{\DH^{\Lambda}(z)}{\DM^{\Lambda}(z)},
\end{equation} where $\DM^{\Lambda}(z)=\int_0^z \DH^{\Lambda}(z')\,{\rm d}z'$ and $\DH^{\Lambda}(z)$ is given by Eq.~(\ref{eq:DHLCDM}). In practice we evaluate $R_{\rm HM}(z)$ at the BAO redshift-bin centers $z_i$ as
$R_{\rm HM}(z_i)=\DH(z_i)/\DM(z_i)$, constructed directly from the anisotropic BAO constraints
$(\DH/\rd,\DM/\rd)$ reported in the same redshift bin. This choice enables us to propagate the full DESI BAO covariance (including the correlation between $\DH$ and $\DM$ within each bin) into the derived $R_{\rm HM}$ and $\OmL$ constraints.

Although one can define a more general hybrid family
\begin{equation}
   R_{\rm HM}(z_i,z_j)\equiv \frac{\DH(z_i)}{\DM(z_j)},
\end{equation} we do not pursue the full two-redshift set in this work. Once the radial--radial ratios
$R_{\rm HH}(z_i,z_j)=\DH(z_i)/\DH(z_j)$ are included, allowing $z_i\neq z_j$ in $R_{\rm HM}$
produces an overcomplete collection of tests that repeatedly reuses the same $\DH(z_i)$ measurements,
leading to strong correlations among the inferred $\OmL$ values and complicating interpretation.
We therefore restrict to the diagonal ratio in Eq.~(\ref{eq:RHM}), which cleanly incorporates transverse information
without duplicating the cross-redshift leverage already captured by $R_{\rm HH}$.

\rev{This restriction is a choice of presentation rather than a claim that all non-diagonal ratios are useless. One could design a smaller set of off-diagonal $R_{\rm HM}(z_i,z_j)$ ratios, for example by choosing widely separated bins or by optimizing an approximately independent subset using the full derived covariance matrix. However, small off-diagonal entries in the published BAO covariance do not by themselves guarantee independence of the derived diagnostics, because the same $\DH(z_i)$ measurements would still be reused and the integrated quantity $\DM(z_j)$ encodes line-of-sight information from lower redshifts. In this paper we therefore avoid interpreting $R_{\rm HH}$, $R_{\rm HM}$, and $R_{\rm HV}$ as fully independent data sets; they are complementary diagnostic families constructed from a shared BAO distance vector.}

Because $\DM^{\Lambda}(z)$ involves an integral, there is no closed-form inversion analogous to Eq.~(\ref{eq:OmL}).
We therefore determine $\OmL$ by numerically solving the one-dimensional equation
$R_{\rm HM}^{\Lambda}(z;\OmL)=R_{\rm HM}(z)$ for $\OmL\in[0,1]$.

\subsubsection{Radial--isotropic test ($R_{\rm HV}$)}
Similarly, we define a ratio between a radial distance and the volume-averaged distance:
\begin{equation}\label{eq:RHV}
   \frac{\DH(z_i)}{\DV(z_j)}\equiv R_{\rm HV}(z_i,z_j)
   \xrightarrow{\Lambda}
   R_{\rm HV}^{\Lambda}(z_i,z_j)\equiv \frac{\DH^{\Lambda}(z_i)}{\DV^{\Lambda}(z_j)},
\end{equation}
where $\DV^{\Lambda}(z)=\left[z\,(\DM^{\Lambda}(z))^2\,\DH^{\Lambda}(z)\right]^{1/3}$.
As for $R_{\rm HM}$, we infer $\OmL$ by numerically solving $R_{\rm HV}^{\Lambda}(z_i,z_j;\OmL)=R_{\rm HV}(z_i,z_j)$.

\subsubsection{\rev{How deviations from flat $\Lambda$CDM appear in the ratio diagnostics}}
\label{sec:deviation_signature}

\begin{figure}
    \centering
    \includegraphics[width=\columnwidth]{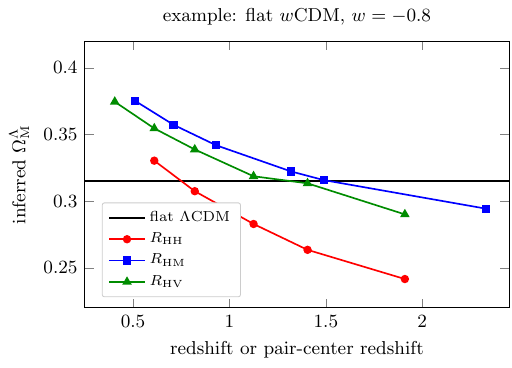}
    \caption{\rev{Illustrative response of the three ratio families to a non-$\Lambda$CDM background. The black horizontal line shows the result expected when noiseless flat-$\Lambda$CDM distances with $\Omega_{\rm M}=0.315$ are analyzed using the same model. The red circles, blue squares, and green triangles use the same color associations as Fig.~\ref{fig:Om} and show the effective $\Omega_{\rm M}^{\Lambda}$ that would be inferred from noiseless flat $w$CDM distances with the same $\Omega_{\rm M}$ but $w=-0.8$, using representative redshifts comparable to the DESI BAO bins. Instead of returning a single constant value, the non-$\Lambda$CDM example produces redshift-dependent and family-dependent inferred $\Omega_{\rm M}^{\Lambda}$ values.}}
    \label{fig:deviation_demo}
\end{figure}

\rev{For any one of the three diagnostic families $X\in\{{\rm HH},{\rm HM},{\rm HV}\}$, a convenient way to visualize a departure from flat $\Lambda$CDM is}
\begin{equation}
    \rev{\Delta_X \equiv R_X^{\rm obs}-R_X^{\Lambda}(\Omega_{\rm ref}), \qquad
    \delta\Omega_{{\rm M},X}^{\Lambda}\equiv \Omega_{{\rm M},X}^{\Lambda}-\Omega_{\rm ref},}
\end{equation}
\rev{where $\Omega_{\rm ref}$ may be chosen, for example, from a best-fit flat-$\Lambda$CDM model or from Planck. In the null case, all $\Delta_X$ vanish (within noise) for a single $\Omega_{\rm ref}$, and all inferred $\Omega_{{\rm M},X}^{\Lambda}$ values are constant in redshift and mutually consistent. A non-$\Lambda$CDM expansion history generally produces a redshift-dependent sequence of $\Omega_{{\rm M},X}^{\Lambda}$ values and may also produce offsets between the radial, radial--transverse, and radial--isotropic families. Figure~\ref{fig:deviation_demo} gives a noiseless illustrative example for a flat $w$CDM model analyzed under the flat-$\Lambda$CDM assumption. The example is not used in the data analysis; it is intended only to show the qualitative signature that the diagnostics are designed to reveal.}

\subsection{Effective-redshift interpretation and redshift ranges}
\label{subsec:zmatch}

\rev{We distinguish two different uses of ``effective redshift.'' The DESI BAO measurements are reported at a tracer-bin effective redshift, conventionally denoted $z_{\rm eff}$, which is determined by the DESI clustering analysis for each tracer sample and redshift bin. In this paper those published DESI values are the redshifts denoted by $z_i$ and $z_j$. By contrast, $z_{\rm c}$, $z_{\rm d}$, and $z_*$ below are mathematical mean-value redshifts introduced only to interpret integrated distances such as $\DM$ and $\DV$ as effective line-of-sight distances over an interval. They are not the same as the DESI $z_{\rm eff}$ values and are not used to redefine the BAO measurement redshifts.}

The transverse and isotropic distances encode \emph{integrated} information about $\DH(z)$.
For interpretation and for visual comparisons, it is useful to associate $\DM$ and $\DV$ with an effective redshift at which they correspond
to a local Hubble distance.
By the integral mean value theorem \citep{MVT}, there exists an intermediate redshift $z_{\rm c}\in(0,z)$ such that
\begin{eqnarray}\label{eq:IMVT}
    \DM(z) &=& z\,\DH(z_{\rm c}), \nonumber\\
    \DM^{\Lambda}(z) &=& z\,\DH^{\Lambda}(z_{\rm c}^{\Lambda}),
\end{eqnarray}
where $z_{\rm c}^{\Lambda}$ is the corresponding intermediate redshift in a given $\Lambda$CDM model.
Likewise, using Eq.~(\ref{eq:DV}) and Eq.~(\ref{eq:IMVT}), one can write
\begin{eqnarray}\label{eq:DVDH}
   \DV(z) &=& \left[z\,\DM^2(z)\,\DH(z)\right]^{1/3} \nonumber\\
          &=& z\left[\DH^2(z_{\rm c})\,\DH(z)\right]^{1/3}
           \equiv z\,\DH(z_{\rm d}),
\end{eqnarray}
for some $z_{\rm d}$ satisfying $z_{\rm c}<z_{\rm d}<z$.

In practice, $z_{\rm c}$ and $z_{\rm d}$ are not directly observable and depend (weakly) on the assumed background model.
To provide a conservative and transparent summary of which redshift interval each ratio probes, we report a \emph{redshift range}
for each test:
for $R_{\rm HH}(z_i,z_j)$ it is simply $[z_i,z_j]$;
for $R_{\rm HM}(z)$ it is $[z_{\rm c},z]$;
and for $R_{\rm HV}(z_i,z_j)$ it is $[z_{\rm d},z_i]$, where $z_{\rm d}<z_j$.
When needed, we bracket these intermediate-redshift mappings by scanning $\OmL$ over a broad prior range (we take $\OmL\in[0,1]$)
to obtain conservative bounds on $z_{\rm c}^{\Lambda}$ and $z_{\rm d}^{\Lambda}$ for quoting the corresponding interval.
Additional details of the mapping are provided in the Appendix.

\subsection{Converting $\DM$ measurements into an effective $\DH$ over a redshift interval}
\label{subsec:DMtoDH}

When $\DM$ is measured at two redshifts $z_1<z_2$, the difference
\begin{eqnarray}
    \label{eq:DM12}
    \DM(z_1,z_2) &\equiv& \DM(z_2)-\DM(z_1) \nonumber\\
                 &=& \int_{z_1}^{z_2}\DH(z')\,{\rm d}z'
                  = (z_2-z_1)\,\DH(z_*),
\end{eqnarray}
for some $z_*\in(z_1,z_2)$ by the integral mean value theorem \citep{MVT}.
This relation defines an \emph{effective} Hubble distance
$\DH(z_*)\equiv \DM(z_1,z_2)/(z_2-z_1)$ associated with the redshift interval $[z_1,z_2]$.
We use $(z_1,z_2)$ to quantify the redshift range for this converted $\DH$ in our figures and tables.

\subsection{Data sets}
\label{subsec:data}

We use BAO measurements from the Dark Energy Spectroscopic Instrument (DESI) Data Release~1 (DR1)
\citep{DESI:2024uvr,DESI:2024lzq} and Data Release~2 (DR2) \citep{DESI:2025zgx}.
The inputs are the published BAO constraints on $\DH/\rd$, $\DM/\rd$, and/or $\DV/\rd$ in each redshift bin, from which we form
the ratios in Eqs.~(\ref{eq:RHH})--(\ref{eq:RHV}).
\rev{These references are the final DESI BAO measurement papers from which we take the published distance constraints, effective redshifts, and covariance matrices. The tracer samples and distance types used in our analysis are summarized in Table~\ref{tab:bao_inputs}.}
Because our diagnostics are ratios of BAO distances, the sound horizon $\rd$ cancels exactly.
Throughout, we use the full DESI BAO covariance matrices provided with the data products, and propagate them into all derived ratios and $\OmL$ constraints.

\begin{table*}
    \centering
    \begin{revblock}
    \small
    \renewcommand{\arraystretch}{1.18}
    \begin{tabular}{l l c l l}
        \hline\hline
        Release & DESI tracer/bin & $z_{\rm eff}$ & Published BAO distance(s) used & Role in this work \\
        \hline
        DR1 & BGS & 0.300 & $\DV/\rd$ & isotropic input; $\DH/\DV$ and converted-$\DH$ tests. \\
        DR1 & LRG1 & 0.510 & $\DM/\rd$, $\DH/\rd$ & anisotropic input; radial and mixed tests. \\
        DR1 & LRG2 & 0.710 & $\DM/\rd$, $\DH/\rd$ & anisotropic input; radial and $\DH/\DM$ tests. \\
        DR1 & LRG3+ELG1 & 0.930 & $\DM/\rd$, $\DH/\rd$ & combined anisotropic input. \\
        DR1 & ELG2 & 1.320 & $\DM/\rd$, $\DH/\rd$ & high-$z$ anisotropic input. \\
        DR1 & QSO & 1.490 & $\DV/\rd$ & isotropic QSO input; $\DH/\DV$ tests. \\
        DR1 & Ly$\alpha$ forest/QSO & 2.330 & $\DM/\rd$, $\DH/\rd$ & high-$z$ Ly$\alpha$ anisotropic input. \\
        \hline
        DR2 & BGS & 0.295 & $\DV/\rd$ & isotropic input; updated DR2 BGS selection. \\
        DR2 & LRG1 & 0.510 & $\DM/\rd$, $\DH/\rd$ & anisotropic input; radial and $\DH/\DM$ tests. \\
        DR2 & LRG2 & 0.706 & $\DM/\rd$, $\DH/\rd$ & anisotropic input; shifted $z_{\rm eff}$. \\
        DR2 & LRG3+ELG1 & 0.934 & $\DM/\rd$, $\DH/\rd$ & combined anisotropic input. \\
        DR2 & ELG2 & 1.321 & $\DM/\rd$, $\DH/\rd$ & high-$z$ anisotropic input. \\
        DR2 & QSO & 1.484 & $\DM/\rd$, $\DH/\rd$ & anisotropic QSO input in DR2. \\
        DR2 & Ly$\alpha$ forest/QSO & 2.330 & $\DM/\rd$, $\DH/\rd$ & high-$z$ Ly$\alpha$ anisotropic input. \\
        \hline\hline
    \end{tabular}
    \caption{Summary of the DESI BAO inputs used in this work. The $z_{\rm eff}$ values are the effective redshifts reported by the DESI BAO analyses for the corresponding tracer bins; they should not be confused with the mean-value redshifts $z_{\rm c}$, $z_{\rm d}$, and $z_*$ introduced in Secs.~\ref{subsec:zmatch}--\ref{subsec:DMtoDH}.}
    \label{tab:bao_inputs}
    \end{revblock}
\end{table*}

\subsection{Uncertainty propagation}
\label{subsec:errors}

We propagate measurement uncertainties using the \emph{full} DESI BAO covariance matrices.
For each data release we construct the BAO data vector (including all reported distance parameters across redshift bins) and its
associated covariance matrix, and generate Monte Carlo realizations by drawing from a multivariate Gaussian distribution.
For each realization we compute the relevant ratio(s) and infer $\OmL$ (analytically for $R_{\rm HH}$ via Eq.~\ref{eq:OmL}, and
numerically for $R_{\rm HM}$ and $R_{\rm HV}$). This procedure automatically propagates all published correlations among the BAO
distance measurements into the derived ratios and $\OmL$ constraints.

\rev{The same Monte Carlo realizations also make clear that many derived ratios are correlated because they share input BAO distances. Consequently, the tables below should be read as diagnostic summaries of individual ratios; they are not a set of statistically independent measurements unless the full derived covariance is used.}

We summarize each derived quantity by the median and central equal-tailed confidence intervals: the 68\% CL is given by the
16th and 84th percentiles, and the 95\% CL by the 2.5th and 97.5th percentiles.

\section{Results}
\label{sec:results}

\begin{figure*}
    \centering
    \includegraphics[width=\textwidth]{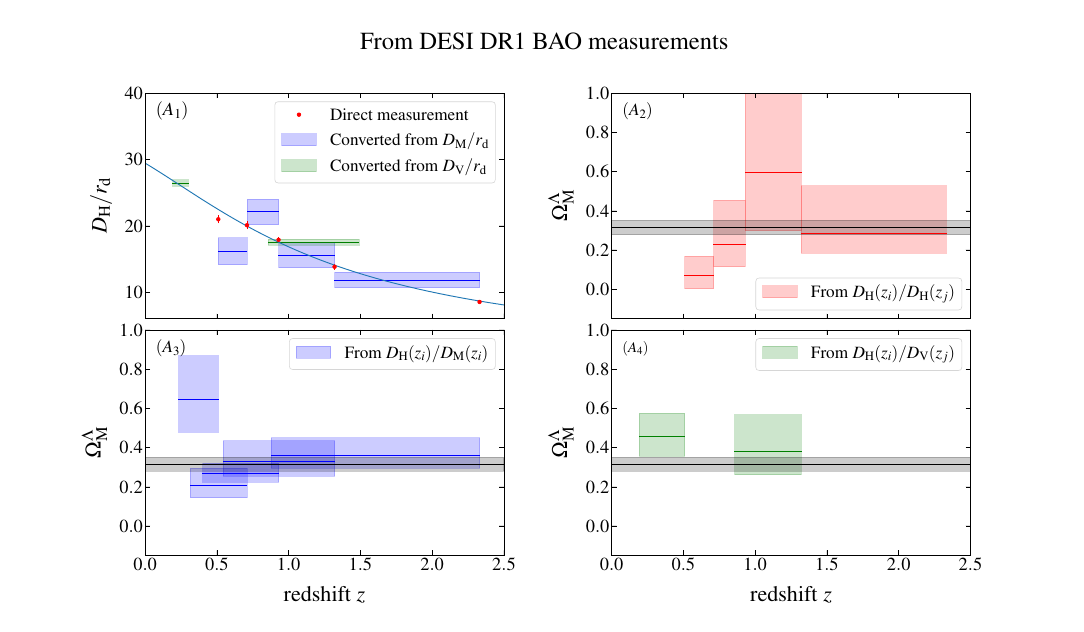}
    \includegraphics[width=\textwidth]{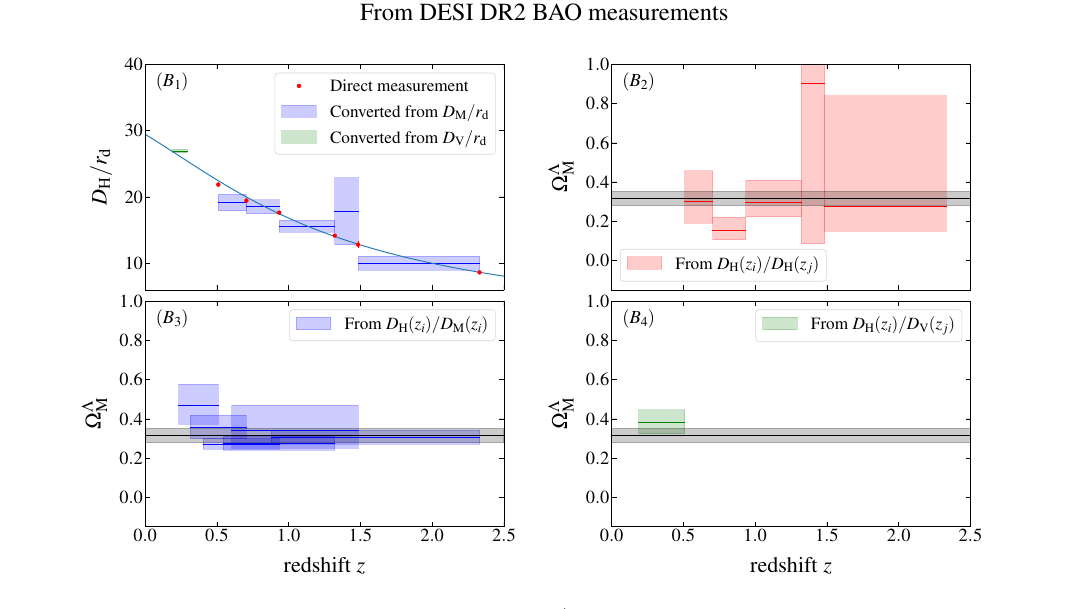}
    \caption{
    \textbf{Summary of BAO distances and ratio-based $\OmL$ null tests.}
    Panel $(A_1)$: $\DH/\rd$ as a function of redshift for DESI DR1, including direct radial BAO measurements (red points),
    and the effective $\DH/\rd$ ranges converted from $\DM/\rd$ (blue bands) and $\DV/\rd$ (green bands) using the redshift-matching
    procedure described in Secs.~\ref{subsec:zmatch}--\ref{subsec:DMtoDH}. Panel $(A_2)$: $\OmL$ inferred from the radial--radial
    ratio $\DH(z_i)/\DH(z_j)$ (red bands). Panel $(A_3)$: $\OmL$ inferred from the radial--transverse ratio $\DH(z_i)/\DM(z_i)$ (blue bands).
    Panel $(A_4)$: $\OmL$ inferred from the radial--isotropic ratio $\DH(z_i)/\DV(z_j)$ (green bands). Panels $(B_1)$--$(B_4)$ show the corresponding results for DESI DR2. The gray band in panels (A2)--(B4) shows the Planck 2018 flat-$\Lambda$CDM constraint on $\Omega_{\rm M}$
(shown for comparison; 68\% CL) \citep{Planck2018}.}
    \label{fig:Om}
\end{figure*}

Figure~\ref{fig:Om} summarizes the BAO distance measurements and the derived null-test constraints on the effective
flat-$\Lambda$CDM matter density parameter $\OmL$ for DESI DR1 and DR2. The corresponding numerical values are listed in
Tables~\ref{tab:Om_DR1} and~\ref{tab:Om_DR2}. Unless stated otherwise, uncertainties quoted in the tables correspond to
central equal-tailed 68\% confidence intervals (the 95\% intervals are defined analogously; see Sec.~\ref{subsec:errors}).

\rev{The color scheme is kept fixed throughout Fig.~\ref{fig:Om}: red denotes quantities based on direct radial $\DH$ measurements or $\DH/\DH$ ratios, blue denotes quantities involving transverse $\DM$ information or $\DH/\DM$ ratios, green denotes quantities involving isotropic $\DV$ information or $\DH/\DV$ ratios, and gray denotes the Planck 2018 flat-$\Lambda$CDM reference band.}

\begin{table*}
    \centering
    \renewcommand{\arraystretch}{1.25}
    \begin{tabular}{c c c c|c c c|c c c c}
        \hline\hline
        \multicolumn{4}{c|}{$\DH(z_i)/\DH(z_j)$} &
        \multicolumn{3}{c|}{$\DH(z_i)/\DM(z_i)$} &
        \multicolumn{4}{c}{$\DH(z_i)/\DV(z_j)$} \\
        \hline
        $z_i$ & $z_j$ & $\OmL$ & $z$-range &
        $z_i$ & $\OmL$ & $z$-range &
        $z_i$ & $z_j$ & $\OmL$ & $z$-range \\
        \hline
        $0.510$ & $0.710$ & $0.069^{+0.098}_{-0.067}$ & $[0.510,0.710]$ &
        $0.510$ & $0.648^{+0.219}_{-0.168}$ & $[0.233,0.510]$ &
        $0.510$ & $0.300$ & $0.454^{+0.122}_{-0.099}$ & $[0.192,0.510]$ \\
        $0.710$ & $0.930$ & $0.228^{+0.226}_{-0.113}$ & $[0.710,0.930]$ &
        $0.710$ & $0.209^{+0.085}_{-0.066}$ & $[0.316,0.710]$ &
        $1.320$ & $1.490$ & $0.382^{+0.185}_{-0.121}$ & $[0.858,1.320]$ \\
        $0.930$ & $1.320$ & $0.598^{+0.402}_{-0.300}$ & $[0.930,1.320]$ &
        $0.930$ & $0.268^{+0.053}_{-0.044}$ & $[0.402,0.930]$ &
        - & - & - & - \\
        $1.320$ & $2.330$ & $0.286^{+0.241}_{-0.103}$ & $[1.320,2.330]$ &
        $1.320$ & $0.330^{+0.104}_{-0.076}$ & $[0.546,1.320]$ &
        - & - & - & - \\
        - & - & - & - &
        $2.330$ & $0.361^{+0.088}_{-0.067}$ & $[0.880,2.330]$ &
        - & - & - & - \\
        \hline\hline
    \end{tabular}
    \caption{
    \textbf{$\OmL$ derived from DESI DR1 BAO measurements.}
    We report $\OmL$ inferred from three ratio tests: radial--radial $\DH(z_i)/\DH(z_j)$, radial--transverse $\DH(z_i)/\DM(z_i)$,
    and radial--isotropic $\DH(z_i)/\DV(z_j)$.
    For ratios involving integrated distances, the quoted $z$-range indicates the effective redshift interval probed by the corresponding
    measurement (Sec.~\ref{subsec:zmatch}).
    Uncertainties are 68\% confidence intervals (Sec.~\ref{subsec:errors}).
    }
    \label{tab:Om_DR1}
\end{table*}

\begin{table*}
    \centering
    \renewcommand{\arraystretch}{1.25}
    \begin{tabular}{c c c c|c c c|c c c c}
        \hline\hline
        \multicolumn{4}{c|}{$\DH(z_i)/\DH(z_j)$} &
        \multicolumn{3}{c|}{$\DH(z_i)/\DM(z_i)$} &
        \multicolumn{4}{c}{$\DH(z_i)/\DV(z_j)$} \\
        \hline
        $z_i$ & $z_j$ & $\OmL$ & $z$-range &
        $z_i$ & $\OmL$ & $z$-range &
        $z_i$ & $z_j$ & $\OmL$ & $z$-range \\
        \hline
        $0.510$ & $0.706$ & $0.298^{+0.159}_{-0.108}$ & $[0.510,0.706]$ &
        $0.510$ & $0.465^{+0.112}_{-0.094}$ & $[0.233,0.510]$ &
        $0.510$ & $0.295$ & $0.381^{+0.064}_{-0.057}$ & $[0.189,0.510]$ \\
        $0.706$ & $0.934$ & $0.153^{+0.067}_{-0.049}$ & $[0.706,0.934]$ &
        $0.706$ & $0.353^{+0.064}_{-0.055}$ & $[0.314,0.706]$ &
        - & - & - & - \\
        $0.934$ & $1.321$ & $0.296^{+0.112}_{-0.075}$ & $[0.934,1.321]$ &
        $0.934$ & $0.271^{+0.028}_{-0.026}$ & $[0.403,0.934]$ &
        - & - & - & - \\
        $1.321$ & $1.484$ & $0.900^{+0.100}_{-0.815}$ & $[1.321,1.484]$ &
        $1.321$ & $0.273^{+0.039}_{-0.033}$ & $[0.546,1.321]$ &
        - & - & - & - \\
        $1.484$ & $2.330$ & $0.277^{+0.564}_{-0.129}$ & $[1.484,2.330]$ &
        $1.484$ & $0.339^{+0.131}_{-0.089}$ & $[0.603,1.484]$ &
        - & - & - & - \\
        - & - & - & - &
        $2.330$ & $0.304^{+0.038}_{-0.033}$ & $[0.880,2.330]$ &
        - & - & - & - \\
        \hline\hline
    \end{tabular}
    \caption{
    \textbf{Same as Table~\ref{tab:Om_DR1} but for DESI DR2 BAO measurements.}
    }
    \label{tab:Om_DR2}
\end{table*}

\subsection{Distance measurements and effective redshift ranges}

Panels $(A_1)$ and $(B_1)$ of Fig.~\ref{fig:Om} compare the directly measured radial BAO distance $\DH/\rd$ (red points) with
effective $\DH/\rd$ ranges converted from transverse $\DM/\rd$ and isotropic $\DV/\rd$ measurements (blue and green bands).
The converted constraints are displayed as bands because they correspond to an effective line-of-sight distance averaged over a finite
redshift interval rather than a strictly local measurement (Sec.~\ref{subsec:zmatch}).
Overall, the direct and converted distances exhibit broad consistency with each other and with the survey best-fit $\Lambda$CDM curve,
providing a basic sanity check of the redshift-matching procedure prior to applying the ratio tests.
\rev{Thus, within the current BAO uncertainties, the radial, transverse, and isotropic DESI distance measurements are mutually compatible for the purposes of this internal consistency test.}

\subsection{Radial--radial ratios: $\DH(z_i)/\DH(z_j)$}

The radial--radial ratio test uses only direct $\DH/\rd$ measurements and is therefore the most direct implementation of our
sound-horizon-independent null test.
The resulting $\OmL$ values are shown in panels $(A_2)$ and $(B_2)$ and listed in the first block of Tables~\ref{tab:Om_DR1}--\ref{tab:Om_DR2}.
For DR1, the inferred $\OmL$ values span a wide range across the available redshift pairs, while remaining statistically consistent with
a constant value given current uncertainties.
For DR2, the additional redshift coverage leads to more ratio combinations, including one high-redshift pair with a highly asymmetric
constraint (Table~\ref{tab:Om_DR2}), reflecting the limited constraining power of that specific ratio at current precision.
\rev{Compared with the Planck 2018 flat-$\Lambda$CDM value shown by the gray band in Fig.~\ref{fig:Om}, several individual radial--radial ratios are broadly compatible while some low- or intermediate-redshift 68\% intervals do not overlap the narrow Planck band. We do not interpret these individual offsets as evidence against flat $\Lambda$CDM, because the derived ratios are correlated and some constraints are highly non-Gaussian; the constant-$\OmL$ summaries in Sec.~\ref{sec:results} show no significant redshift evolution within the current precision.}

\subsection{Radial--transverse ratios: $\DH(z_i)/\DM(z_i)$}

Panels $(A_3)$ and $(B_3)$ show $\OmL$ inferred from ratios combining a local distance ($\DH$) with an integrated distance ($\DM$).
Because $\DM$ integrates $\DH$ over $0<z'<z$, these constraints probe a redshift interval and are presented as bands spanning the
corresponding effective $z$-range (Sec.~\ref{subsec:zmatch}).
Compared to the radial-only ratios, the $\DH/\DM$ diagnostic yields a relatively stable set of $\OmL$ values across redshift in both DR1 and DR2,
with the tightest constraints coming from the mid-redshift bins where $\DM/\rd$ is measured most precisely.
\rev{The mid-redshift $\DH/\DM$ constraints are close to the Planck reference value, while the lowest-redshift bins give somewhat higher inferred $\OmL$. Given their uncertainties and shared covariance with the radial measurements, these offsets remain consistent with the statement that the radial and transverse BAO distances are mutually compatible at present precision.}

\subsection{Radial--isotropic ratios: $\DH(z_i)/\DV(z_j)$}

The $\DH/\DV$ ratio combines the local expansion rate with the volume-averaged BAO distance.
Given the current set of available $\DV$ measurements, the number of independent $\DH/\DV$ ratios is limited (Tables~\ref{tab:Om_DR1}--\ref{tab:Om_DR2}),
and the corresponding $\OmL$ constraints remain comparatively weak.
Nevertheless, the derived values provide an additional cross-check that mixes radial and isotropic BAO information in a manner that remains
independent of the absolute BAO scale.
\rev{The available $\DH/\DV$ values lie modestly above the Planck reference band in Fig.~\ref{fig:Om}, but their uncertainties are broad and the present data provide too few independent mixed radial--isotropic ratios to establish a statistically significant discrepancy.}

\subsection{Summary and DR1--DR2 comparison}

To quantify the internal consistency of each diagnostic family, we perform a simple constant-$\OmL$ fit within each family and data release.
As a descriptive statistic, we treat the entries as independent and use symmetrized 68\% uncertainties; this is intended as a compact summary
rather than a full likelihood analysis (which would require the complete covariance of the derived ratios).
With this convention, the best-fit constants are approximately:
$\OmL\simeq 0.146$ for the DR1 $\DH/\DH$ ratios and $\OmL\simeq 0.214$ for DR2, with $\chi^2/{\rm dof}\simeq 1.14$ in both cases;
for the $\DH/\DM$ ratios we obtain $\OmL\simeq 0.293$ for both DR1 and DR2, with $\chi^2/{\rm dof}\simeq 1.45$ (DR1) and
$\chi^2/{\rm dof}\simeq 1.01$ (DR2).
Within the current precision, these fits show no compelling evidence for redshift evolution \emph{within} each diagnostic family, while the
differences \emph{between} diagnostic families motivate the discussion in Sec.~\ref{sec:conclusion}.

\rev{The small changes in the quoted $z$-ranges between DR1 and DR2 follow from the published DESI effective redshifts and from the set of BAO quantities reported in each release. For example, the BGS isotropic effective redshift changes from $z_{\rm eff}=0.300$ in DR1 to $0.295$ in DR2, the LRG2 and LRG3+ELG1 effective redshifts shift slightly, and the QSO measurement changes from an isotropic $\DV/\rd$ input in DR1 to anisotropic $\DM/\rd$ and $\DH/\rd$ inputs in DR2. These changes arise from the DESI sample definitions, redshift calibration, and BAO fitting choices in the two releases; they are not introduced by our mean-value redshift mapping.}

\section{Conclusion}
\label{sec:conclusion}

We have presented a simple BAO-based null test of spatially flat $\Lambda$CDM built from ratios of BAO distance measurements.
The central idea is to map each observed ratio to an \emph{effective} flat-$\Lambda$CDM matter density parameter, $\OmL$,
defined as the value of $\Omega_{\rm M}$ for which the corresponding $\Lambda$CDM prediction reproduces the measured ratio.
Because BAO constraints are typically reported in units of the sound horizon, and because our diagnostics use ratios of BAO
distances, the dependence on the absolute BAO scale $\rd$ cancels identically; moreover, the overall normalization set by
$H_0$ drops out of the ratios. This yields a transparent, calibration-free internal consistency check that can combine radial
and transverse BAO information within a single framework.

We applied the test to BAO measurements from the DESI Data Release~1 and
Data Release~2 \citep{DESI:2024uvr,DESI:2025zgx}, using three complementary families of ratios:
a purely radial test based on $\DH(z_i)/\DH(z_j)$, and two mixed tests combining $\DH$ with the integrated distances $\DM$
and $\DV$. For transverse and isotropic BAO, we adopted a redshift-matching strategy based on the integral mean value theorem
to associate integrated distances with an effective line-of-sight distance over a well-defined redshift interval.
All inferred ratios and $\OmL$ constraints were obtained by propagating the \emph{full} DESI BAO covariance matrices through
Monte Carlo sampling. Within current uncertainties, the inferred $\OmL$ values are broadly consistent with a redshift-independent
constant, and hence with the flat-$\Lambda$CDM expectation. Mild deviations appear for some redshift ranges in the mixed
$\DH/\DM$ and $\DH/\DV$ diagnostics, but these are not statistically significant at the present precision.

Interpreted as a null test, any statistically significant redshift dependence of $\OmL$ \emph{within} a given diagnostic family,
or a robust inconsistency \emph{between} different families constructed from the same dataset, would signal a departure from
flat $\Lambda$CDM. Such a failure could arise from dynamical dark energy, spatial curvature, modified gravity, or residual
observational systematics, and would therefore motivate further investigation rather than a unique physical interpretation.
The ratio-based nature of the test makes it particularly useful for tracking internal consistency as BAO measurements improve.

Looking ahead, future DESI data releases, together with forthcoming wide-area surveys---including the Subaru Prime Focus
Spectrograph (PFS) \citep{Takada2014}, ESA's \textit{Euclid} mission \citep{Laureijs2011}, and the Nancy Grace Roman Space
Telescope \citep{Spergel2015,Akeson2019}---will provide more precise distance measurements over wider redshift ranges,
enabling substantially more stringent tests of the constancy of $\OmL$ and allowing finer redshift binning and a larger set
of ratio combinations. The same framework can also be generalized to additional background models (e.g.\ including curvature
or simple $w$CDM extensions), and can be combined with independent late-time probes to help diagnose the origin of any detected
inconsistencies.

\section*{Acknowledgements}
We thank Gong-Bo Zhao for insightful discussions. Z.C. acknowledges support from National Key R\&D Program of China (grant no. 2023YFA1605600), National Natural Science  Foundation of China (\#12525303), and Tsinghua University Initiative Scientific Research Program.

\begin{revblock}
This research used data obtained with the Dark Energy Spectroscopic Instrument (DESI). DESI construction and operations is managed by the Lawrence Berkeley National Laboratory. This material is based upon work supported by the U.S. Department of Energy, Office of Science, Office of High-Energy Physics, under Contract No. DE--AC02--05CH11231, and by the National Energy Research Scientific Computing Center, a DOE Office of Science User Facility under the same contract. Additional support for DESI was provided by the U.S. National Science Foundation (NSF), Division of Astronomical Sciences under Contract No. AST-0950945 to the NSF's National Optical-Infrared Astronomy Research Laboratory; the Science and Technology Facilities Council of the United Kingdom; the Gordon and Betty Moore Foundation; the Heising-Simons Foundation; the French Alternative Energies and Atomic Energy Commission (CEA); the National Council of Humanities, Science and Technology of Mexico (CONAHCYT); the Ministry of Science and Innovation of Spain (MICINN), and by the DESI Member Institutions: \url{https://www.desi.lbl.gov/collaborating-institutions}. The DESI collaboration is honored to be permitted to conduct scientific research on I'oligam Du'ag (Kitt Peak), a mountain with particular significance to the Tohono O'odham Nation. Any opinions, findings, and conclusions or recommendations expressed in this material are those of the author(s) and do not necessarily reflect the views of the U.S. National Science Foundation, the U.S. Department of Energy, or any of the listed funding agencies.
\end{revblock}

\section*{Data Availability}
The BAO measurements used in this work are taken from the DESI public data releases \citep{DESI:2024uvr,DESI:2025zgx}.
\rev{Because this article uses public DESI BAO data products, we include the DESI public-data acknowledgement in the Acknowledgements. No DESI data products are redistributed in this article.}
The analysis code (written in Julia) used to produce the results and figures is available at \url{https://github.com/adamcosmology/newOm_BAO/}.
\rev{A frozen archival release of the code is available in a DOI-issuing repository (\url{10.5281/zenodo.19976910}).}

\newpage

\appendix

\section{Distance mapping for $\DM$ and $\DV$}
\label{app:zmap}

Transverse and isotropic BAO distances encode \emph{integrated} information about the local Hubble distance
$\DH(z)\equiv c/H(z)$. For the purposes of interpretation and for quoting the redshift intervals probed by
ratios involving $\DM$ or $\DV$, it is useful to associate these integrated distances with an ``effective''
line-of-sight distance at a lower redshift.

\subsection{Mapping $\DM(z)$ to an effective $\DH(z_{\rm c})$}

By definition,
\begin{equation}
  \DM(z)=\int_0^z \DH(z')\,{\rm d}z'.
\end{equation}
Since $\DH(z)$ is continuous in any standard cosmological model, the integral mean value theorem \citep{MVT}
guarantees the existence of an intermediate redshift $z_{\rm c}\in(0,z)$ such that
\begin{equation}
  \DM(z)=z\,\DH(z_{\rm c})
  \quad\Longleftrightarrow\quad
  \DH(z_{\rm c})=\frac{\DM(z)}{z}.
\end{equation}
In a given background model (e.g.\ flat $\Lambda$CDM specified by $\Om$), the mapping $z\mapsto z_{\rm c}(z)$
can be obtained by numerically solving $\DH(z_{\rm c})=\DM(z)/z$.

\subsection{Mapping $\DV(z)$ to an effective $\DH(z_{\rm d})$}

The isotropic BAO distance is
\begin{equation}
  \DV(z)=\left[z\,\DM^2(z)\,\DH(z)\right]^{1/3}.
\end{equation}
Combining this with $\DM(z)=z\,\DH(z_{\rm c})$ yields
\begin{equation}
  \DV(z)=z\left[\DH^2(z_{\rm c})\,\DH(z)\right]^{1/3}\equiv z\,\DH(z_{\rm d}),
\end{equation}
for some $z_{\rm d}$ that satisfies $z_{\rm c}<z_{\rm d}<z$ \citep{MVT}. Equivalently,
\begin{equation}
  \DH(z_{\rm d})=\frac{\DV(z)}{z},
  \qquad z_{\rm d}\in(z_{\rm c},z),
\end{equation}
and, for a specified background model, $z_{\rm d}$ is obtained by numerically solving the above equation.

\subsection{Conservative bracketing and the quoted $z$-ranges}
\label{app:zrange}

The intermediate redshifts $z_{\rm c}$ and $z_{\rm d}$ are not directly observable, and the mapping depends (weakly)
on the background model through $\DH(z)$. In this work we use these mappings only to provide a transparent \emph{redshift range}
for ratios involving integrated distances (see Sec.~\ref{subsec:zmatch}). In flat $\Lambda$CDM, both $z_{\rm c}(z)$ and $z_{\rm d}(z)$
decrease monotonically as $\Om$ increases; therefore, choosing $\Om=1$ yields a conservative \emph{lower bound} on the effective redshift
associated with a given $\DM(z)$ or $\DV(z)$. To bracket the mapping, we scan $\Om$ over a broad prior $\Om\in[0,1]$ and report the
corresponding range of $z_{\rm c}$ (or $z_{\rm d}$) when quoting the $z$-interval probed by $\DH/\DM$ and $\DH/\DV$ ratios.

\begin{figure*}
  \centering
  \includegraphics[width=\textwidth]{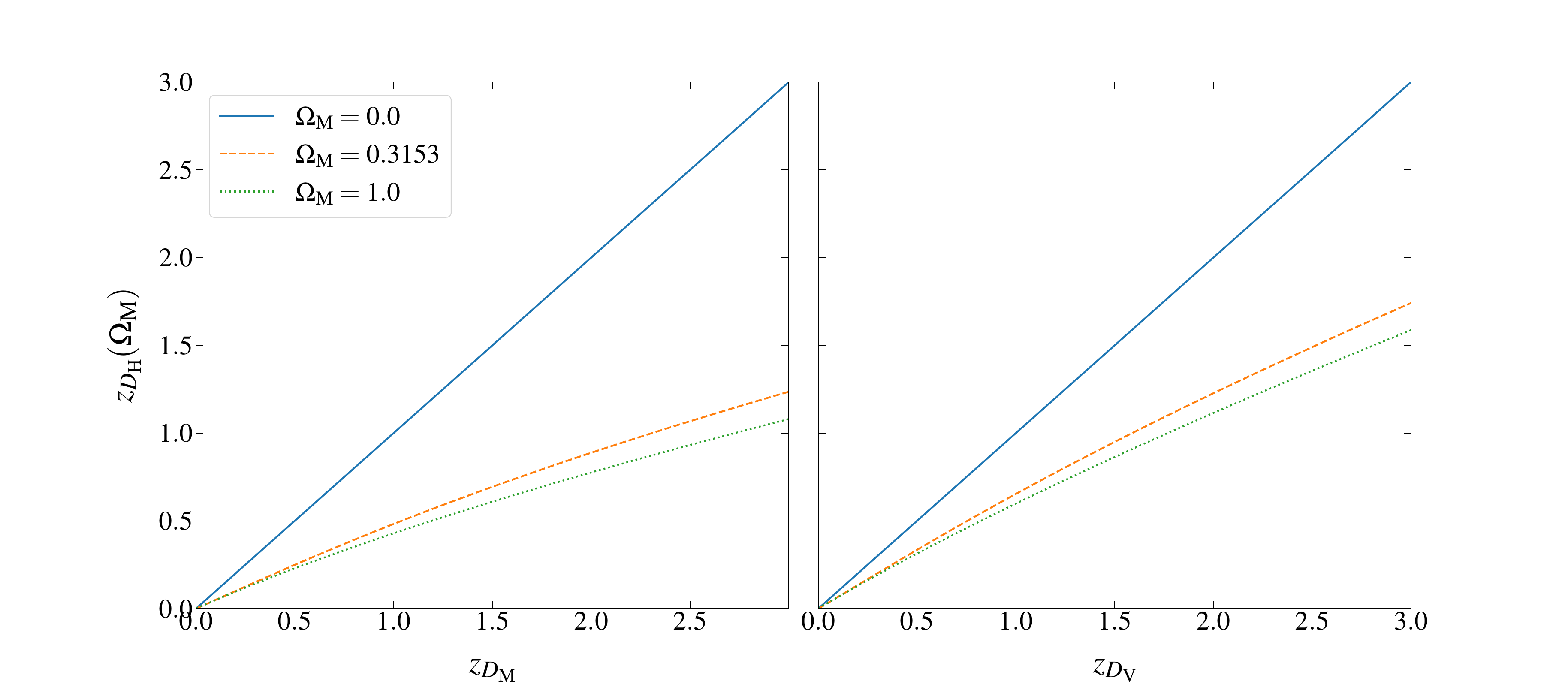}
  \caption{Effective-redshift mapping implied by $\DM(z)=z\,\DH(z_{\rm c})$ (left) and $\DV(z)=z\,\DH(z_{\rm d})$ (right) in flat
  $\Lambda$CDM for representative values of $\Om$. For each $z$, the mapped redshifts satisfy $0<z_{\rm c}<z_{\rm d}<z$.
  Increasing $\Om$ shifts both $z_{\rm c}$ and $z_{\rm d}$ to lower values, so $\Om=1$ provides a conservative lower bound for the
  effective redshift associated with a given $\DM(z)$ or $\DV(z)$.}
  \label{fig:zmap}
\end{figure*}

Figure~\ref{fig:zmap} illustrates the mapping between the redshift at which $\DM$ or $\DV$ is measured and the effective redshift at which
a local $\DH$ reproduces $\DM(z)/z$ or $\DV(z)/z$ in flat $\Lambda$CDM for representative values of $\Om$.

\bibliography{draft}{}

@article{Riess1998,
  author = {Riess, A. G. and Filippenko, A. V. and Challis, P. and others},
  title = {Observational Evidence from Supernovae for an Accelerating Universe and a Cosmological Constant},
  journal = {AJ},
  volume = {116},
  pages = {1009},
  year = {1998},
  doi = {10.1086/300499}
}

@article{Perlmutter1999,
  author = {Perlmutter, S. and Aldering, G. and Goldhaber, G. and others},
  title = {Measurements of Omega and Lambda from 42 High-Redshift Supernovae},
  journal = {ApJ},
  volume = {517},
  pages = {565},
  year = {1999},
  doi = {10.1086/307221}
}

@article{Weinberg1989,
  author = {Weinberg, S.},
  title = {The Cosmological Constant Problem},
  journal = {Rev. Mod. Phys.},
  volume = {61},
  pages = {1},
  year = {1989},
  doi = {10.1103/RevModPhys.61.1}
}

@article{Carroll2001,
  author = {Carroll, S. M.},
  title = {The Cosmological Constant},
  journal = {Living Rev. Relativ.},
  volume = {4},
  pages = {1},
  year = {2001},
  doi = {10.12942/lrr-2001-1}
}

@article{Ratra1988,
  author = {Ratra, B. and Peebles, P. J. E.},
  title = {Cosmological Consequences of a Rolling Homogeneous Scalar Field},
  journal = {Phys. Rev. D},
  volume = {37},
  pages = {3406},
  year = {1988},
  doi = {10.1103/PhysRevD.37.3406}
}

@article{Caldwell1998,
  author = {Caldwell, R. R. and Dave, R. and Steinhardt, P. J.},
  title = {Cosmological Imprint of an Energy Component with General Equation of State},
  journal = {Phys. Rev. Lett.},
  volume = {80},
  pages = {1582},
  year = {1998},
  doi = {10.1103/PhysRevLett.80.1582}
}

@article{Smoot1992,
  author = {Smoot, G. F. and Bennett, C. L. and Kogut, A. and Wright, E. L. and others},
  title = {Structure in the COBE Differential Microwave Radiometer First-Year Maps},
  journal = {ApJL},
  volume = {396},
  pages = {L1},
  year = {1992},
  doi = {10.1086/186504}
}

@article{Bennett1996,
  author = {Bennett, C. L. and Banday, A. J. and Gorski, K. M. and others},
  title = {Four-Year COBE DMR Cosmic Microwave Background Observations},
  journal = {ApJL},
  volume = {464},
  pages = {L1},
  year = {1996},
  doi = {10.1086/310075}
}

@article{deBernardis2000,
  author = {de Bernardis, P. and Ade, P. and Bock, J. and others},
  title = {A Flat Universe from High-Resolution Maps of the Cosmic Microwave Background Radiation},
  journal = {Nature},
  volume = {404},
  pages = {955},
  year = {2000},
  doi = {10.1038/35010035}
}

@article{Halverson2002,
  author = {Halverson, N. W. and Leitch, E. M. and Pryke, C. and others},
  title = {DASI First Results: A Measurement of the Cosmic Microwave Background Angular Power Spectrum},
  journal = {ApJ},
  volume = {568},
  pages = {38},
  year = {2002},
  doi = {10.1086/338879}
}

@book{Peebles1980,
  author = {Peebles, P. J. E.},
  title = {The Large-Scale Structure of the Universe},
  publisher = {Princeton University Press},
  address = {Princeton, NJ},
  year = {1980}
}

@article{Eisenstein2005,
  author = {Eisenstein, D. J. and Zehavi, I. and Hogg, D. W. and others},
  title = {Detection of the Baryon Acoustic Peak in the Large-Scale Correlation Function of SDSS Luminous Red Galaxies},
  journal = {ApJ},
  volume = {633},
  pages = {560},
  year = {2005},
  doi = {10.1086/466512}
}

@article{Cole2005,
  author = {Cole, S. and Percival, W. J. and Peacock, J. A. and others},
  title = {The 2dF Galaxy Redshift Survey: Power-Spectrum Analysis of the Final Dataset and Cosmological Implications},
  journal = {MNRAS},
  volume = {362},
  pages = {505},
  year = {2005},
  doi = {10.1111/j.1365-2966.2005.09318.x}
}

@article{Alam2021,
  author = {Alam, S. and Aubert, M. and Avila, S. and others},
  title = {Completed SDSS-IV Extended Baryon Oscillation Spectroscopic Survey: Cosmological Implications from Two Decades of Spectroscopic Surveys at the Apache Point Observatory},
  journal = {Phys. Rev. D},
  volume = {103},
  pages = {083533},
  year = {2021},
  doi = {10.1103/PhysRevD.103.083533}
}

@article{DESI:2024uvr,
  author = {{DESI Collaboration} and Adame, A. G. and others},
  title = {DESI 2024 III: Baryon Acoustic Oscillations from Galaxies and Quasars},
  journal = {JCAP},
  volume = {04},
  pages = {012},
  year = {2025},
  doi = {10.1088/1475-7516/2025/04/012}
}

@article{DESI:2024lzq,
  author = {{DESI Collaboration} and Adame, A. G. and others},
  title = {DESI 2024 IV: Baryon Acoustic Oscillations from the Lyman Alpha Forest},
  journal = {JCAP},
  volume = {01},
  pages = {124},
  year = {2025},
  doi = {10.1088/1475-7516/2025/01/124}
}

@article{DESI:2025zgx,
  author = {{DESI Collaboration} and Abdul Karim, M. and others},
  title = {DESI DR2 Results II: Measurements of Baryon Acoustic Oscillations and Cosmological Constraints},
  journal = {Phys. Rev. D},
  volume = {112},
  pages = {083515},
  year = {2025},
  doi = {10.1103/tr6y-kpc6}
}

@article{Sahni2008,
  author = {Sahni, V. and Shafieloo, A. and Starobinsky, A. A.},
  title = {Two New Diagnostics of Dark Energy},
  journal = {Phys. Rev. D},
  volume = {78},
  pages = {103502},
  year = {2008},
  doi = {10.1103/PhysRevD.78.103502}
}

@article{Zunckel2008,
  author = {Zunckel, C. and Clarkson, C.},
  title = {Consistency Tests for the Cosmological Constant},
  journal = {Phys. Rev. Lett.},
  volume = {101},
  pages = {181301},
  year = {2008},
  doi = {10.1103/PhysRevLett.101.181301}
}

@article{Sahni2014,
  author = {Sahni, V. and Shafieloo, A. and Starobinsky, A. A.},
  title = {Model Independent Evidence for Dark Energy Evolution from Baryon Acoustic Oscillations},
  journal = {ApJL},
  volume = {793},
  pages = {L40},
  year = {2014},
  doi = {10.1088/2041-8205/793/2/L40}
}

@article{Shafieloo2012,
  author = {Shafieloo, A. and Sahni, V. and Starobinsky, A. A.},
  title = {Is Cosmic Acceleration Slowing Down?},
  journal = {Phys. Rev. D},
  volume = {86},
  pages = {103527},
  year = {2012},
  doi = {10.1103/PhysRevD.86.103527}
}

@book{MVT,
  author = {Rudin, W.},
  title = {Principles of Mathematical Analysis},
  edition = {3},
  publisher = {McGraw-Hill},
  address = {New York},
  year = {1976}
}

@article{Planck2018,
  author = {{Planck Collaboration} and Aghanim, N. and Akrami, Y. and Ashdown, M. and others},
  title = {Planck 2018 Results. VI. Cosmological Parameters},
  journal = {A\&A},
  volume = {641},
  pages = {A6},
  year = {2020},
  doi = {10.1051/0004-6361/201833910}
}

@article{Takada2014,
  author = {Takada, M. and others},
  title = {Extragalactic Science, Cosmology, and Galactic Archaeology with the Subaru Prime Focus Spectrograph},
  journal = {PASJ},
  volume = {66},
  pages = {R1},
  year = {2014},
  doi = {10.1093/pasj/pst019}
}

@article{Laureijs2011,
  author = {Laureijs, R. and Amiaux, J. and Arduini, S. and others},
  title = {Euclid Definition Study Report},
  journal = {arXiv e-prints},
  year = {2011},
  doi = {10.48550/arXiv.1110.3193}
}

@article{Spergel2015,
  author = {Spergel, D. and Gehrels, N. and Baltay, C. and others},
  title = {Wide-Field InfrarRed Survey Telescope-Astrophysics Focused Telescope Assets WFIRST-AFTA 2015 Report},
  journal = {arXiv e-prints},
  year = {2015},
  doi = {10.48550/arXiv.1503.03757}
}

@article{Akeson2019,
  author = {Akeson, R. and others},
  title = {The Wide Field Infrared Survey Telescope: 100 Hubbles for the 2020s},
  journal = {arXiv e-prints},
  year = {2019},
  doi = {10.48550/arXiv.1902.05569}
}
\bibliographystyle{aasjournal}

\end{document}